\documentclass[a4paper,amsmath,amssymb,prb,preprintnumbers,superscriptaddress,twocolumn,showpacs,longbibliography]{revtex4-1} 

\usepackage{graphicx}			
\usepackage{dcolumn}			
\usepackage{bm}					
\usepackage{amssymb}
\usepackage{epsfig}
\usepackage{gensymb}
\usepackage{multirow}
\usepackage{array}
\usepackage{ctable}

\newcommand{\lapprox}{{\scriptscriptstyle\stackrel{<}{\sim}}}

\newif\ifcom
\newif\ifdel

\comtrue               			%
\deltrue          			      %

\begin{document}

\title{NanoSQUID magnetometry of individual cobalt nanoparticles grown by focused electron beam induced deposition}

\author{M. J. Mart\'inez-P\'erez}
\email{mariajose.martinez@uni-tuebingen.de \\ pepamartinez@gmail.com}
\affiliation{Physikalisches Institut -- Experimentalphysik II and Center for Quantum Science (CQ) in LISA${^+}$,
Universit\"at T\"ubingen, Auf der Morgenstelle 14, D-72076 T\"ubingen, Germany}

\author{B. M\"uller}
\affiliation{Physikalisches Institut -- Experimentalphysik II and Center for Quantum Science (CQ) in LISA${^+}$,
Universit\"at T\"ubingen, Auf der Morgenstelle 14, D-72076 T\"ubingen, Germany}

\author{D. Schwebius}
\affiliation{Physikalisches Institut -- Experimentalphysik II and Center for Quantum Science (CQ) in LISA${^+}$,
Universit\"at T\"ubingen, Auf der Morgenstelle 14, D-72076 T\"ubingen, Germany}

\author{D. Korinski}
\affiliation{Physikalisches Institut -- Experimentalphysik II and Center for Quantum Science (CQ) in LISA${^+}$,
Universit\"at T\"ubingen, Auf der Morgenstelle 14, D-72076 T\"ubingen, Germany}

\author{R. Kleiner}
\affiliation{Physikalisches Institut -- Experimentalphysik II and Center for Quantum Science (CQ) in LISA${^+}$,
Universit\"at T\"ubingen, Auf der Morgenstelle 14, D-72076 T\"ubingen, Germany}

\author{J. Ses\'e}
\affiliation{Laboratorio de Microscop\'ias Avanzadas (LMA), Instituto de Nanociencia de Arag\'on (INA), Universidad de Zaragoza, E-50018 Zaragoza, Spain}

\author{D. Koelle}
\affiliation{Physikalisches Institut -- Experimentalphysik II and Center for Quantum Science (CQ) in LISA${^+}$,
Universit\"at T\"ubingen, Auf der Morgenstelle 14, D-72076 T\"ubingen, Germany}


\begin{abstract}
We demonstrate the operation of low-noise nano superconducting quantum interference devices (SQUIDs) based on the high critical field and high critical temperature superconductor YBa$_2$Cu$_3$O$_7$ (YBCO) as ultra-sensitive magnetometers for single magnetic nanoparticles (MNPs). 
The nanoSQUIDs exploit the Josephson behavior of YBCO grain boundaries and have been patterned by focused ion beam milling. 
This allows to precisely define the lateral dimensions of the SQUIDs so as to achieve large magnetic coupling between the nanoloop and individual MNPs.
By means of focused electron beam induced deposition, cobalt MNPs with typical size of several tens of nm have been grown directly on the surface of the sensors with nanometric spatial resolution.
Remarkably, the nanoSQUIDs are operative over extremely broad ranges of applied magnetic field (--1\,T $< \mu_0 H <$ 1\,T) and temperature (0.3\,K $< T<$ 80\,K).
All these features together have allowed us to perform magnetization measurements under different ambient conditions and to detect the magnetization reversal of individual Co MNPs with magnetic moments (1  -- 30) $\times 10^6\,\mu_{\rm B}$.
Depending on the dimensions and shape of the particles we have distinguished between two different magnetic states yielding different reversal mechanisms.
The magnetization reversal is thermally activated over an energy barrier, which has been quantified for the (quasi) single-domain particles.
Our measurements serve to show not only the high sensitivity achievable with YBCO nanoSQUIDs, but also demonstrate that these sensors are exceptional magnetometers for the investigation of the properties of individual nanomagnets.
\end{abstract}

\pacs{}


\maketitle


\section{Introduction}
\label{sec:introduction}

Magnetic nanoparticles (MNPs) are targeted by the scientific community and industry.
After recognizing the large number of size and shape-dependent properties of MNPs, a huge range of potential applications became immediately evident.
Just to mention a few, these properties include magnetic anisotropy (memory),\cite{Morup94,Luis02} phase transitions,\cite{Landau76} magnetocaloric effects\cite{Evangelisti10} or resonance frequencies.\cite{Guslienko02}
Very different industrial sectors have already benefited from the use of MNPs  starting from electronics and information technologies up to medical diagnostics and cancer therapy.\cite{MRSBull13}
In addition, fundamental research on MNPs might also find applications in solid-state quantum information technologies\cite{Leuenberger01} and molecular spintronics.\cite{Bogani08}
In this regard, developing tools for magnetic characterization of small amounts of MNPs or, if possible, individual ones, represents an important step towards the realization and fine-tuning of the properties of MNPs for different applications.

Experiments on individual MNPs were pioneered by Wernsdorfer  and collaborators (for reviews see, e.g., Refs.~[\onlinecite{Wernsdorfer01},\onlinecite{Wernsdorfer09}]).
Among a vast amount of studies, this group succeeded in demonstrating experimentally, e.g., magnetization reversal as described by the N\'eel-Brown\cite{Wernsdorfer97a,Neel49,Brown63} and Stoner-Wohlfarth model\cite{Bonet99,Stoner48} or the occurrence of macroscopic quantum tunneling of the vector magnetic moment.\cite{Wernsdorfer97}
The magnetometers used for this goal were microscopic superconducting quantum interference devices (SQUIDs) based on niobium  thin films.
Since then, SQUID sensors have been further miniaturized to the nanosocopic scale, boosting enormously their sensitivity and noise performance.\cite{Granata16,Martinez-Perez16a}
However, the realization of routine magnetization measurements and the investigation of interesting physics using nanoSQUIDs is still quite limited.\cite{Granata16,Martinez-Perez16a}

Among other reasons, this lack is mainly due to (i) restrictions imposed on the SQUID operation ranges of applied magnetic field $\bm H$ and temperature $T$, which are often much smaller than what is usually required for comprehensive characterization of magnetic materials and (ii) the difficulty of positioning individual MNPs with high spatial precision close to the nanoSQUID loop, which is crucial to achieve the sensitivity required to detect the tiny magnetic moment of MNPs.

We have overcome the first mentioned challenge by using recently developed ultra-sensitive nanoSQUID sensors based on the high critical field and high critical temperature ($T_{\rm c}$)  superconductor YBa$_2$Cu$_3$O$_7$ (YBCO) and submicron grain boundary Josephson junctions.\cite{Nagel11,Schwarz13,Schwarz15}
This approach allows sensor operation at remarkably large in-plane applied magnetic fields (up to one Tesla) and a large range of temperatures (300\,mK -- 80\,K).
Regarding the second issue, cobalt MNPs have been directly grown at precise positions by focused electron beam induced deposition (FEBID).\cite{DeTeresa16}
Being polycrystalline, FEBID-Co is a soft magnetic material with negligible volume-averaged magnetocrystalline anisotropy. 
The equilibrium magnetic state of these MNPs will, therefore, result from the competition between the exchange and magnetotstatic (shape) energies.\cite{Fernandez-Pacheco09}
The use of FEBID allows us to control not only the particle location with nanometric resolution, but also its size and shape.
This gives access to investigating the boundary between single-domain MNPs (dominated by the minimization of the exchange energy) and more complicate spin configurations of topological origin (dominated by the minimization of the magnetostatic energy).\cite{Guslienko08,Fruchart05}

Here, we present nanoSQUID magnetization measurements on five different FEBID-Co MNPs by using a set of five nanoSQUIDs SQ$\# i$ containing MNPs labeled as $\# i$, with $i=1 - 5$, respectively.
NanoSQUID fabrication, operation and electrical characterization is presented in Sec.~\ref{sec:characterization} along with the calculation of their corresponding position-dependent magnetic coupling and spin sensitivity.
MNP growth is described in Sec.~\ref{sec:MNP-patterning}, followed by the description of the magnetization measurements in Sec.~\ref{sec:magnetization-measurements}.
Within this section the total magnetic moment per particle is estimated and the temperature and angular dependence of the switching magnetic fields is analyzed in detail.
Section \ref{sec:conclusions} is left for conclusions.

\section{NanoSQUID characterization}
\label{sec:characterization}

\subsection{NanoSQUID fabrication}
\label{subsec:fabrication}

\begin{figure}[t]
\includegraphics[width=\columnwidth]{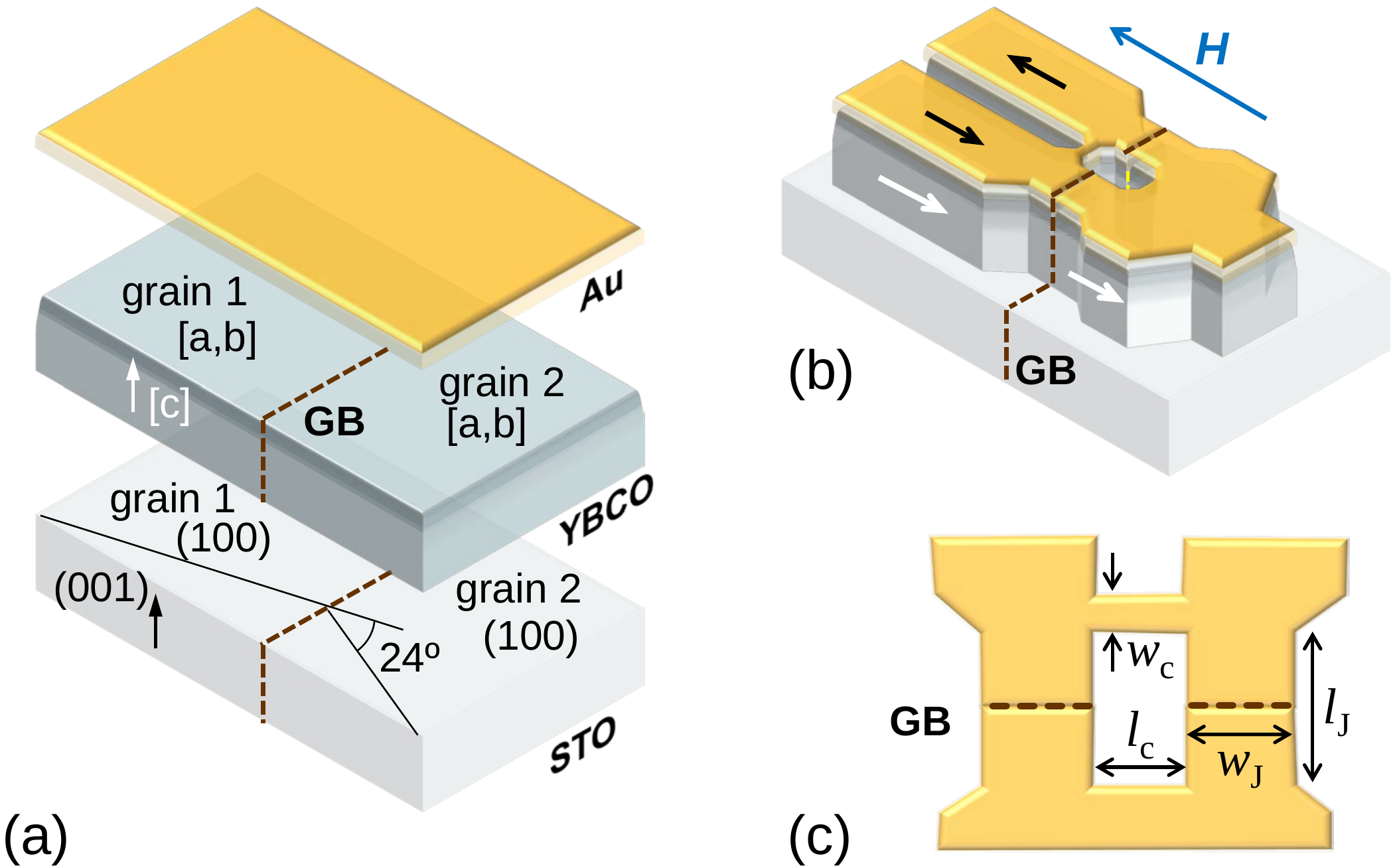}
\caption{YBCO nanoSQUID fabrication:
(a) Scheme of the thin-film deposition process: a 120\,nm-thick YBCO film is grown epitaxially on a STO bicrystal substrate leading to the natural formation of a grain boundary (GB; dashed line). 
The YBCO is covered with a 70\,nm-thick Au layer. 
(b) Final layout of the device after FIB patterning. 
White and black arrows indicate direction of bias and modulation current, respectively.
An external magnetic field $H$ (blue arrow) can be applied in the plane of the SQUID loop.
(c) Schematic top view of the SQUID loop, indicating the relavant geometric parameters.}
\label{Fig1}
\end{figure}

The fabrication of the devices is summarized in Fig.~\ref{Fig1} and briefly described in the following (see Ref.~[\onlinecite{Schwarz13}] for further details).
A 120\,nm thick YBCO film is grown epitaxially by pulsed laser deposition on a SrTiO$_3$ (STO) bicrystal substrate, leading to the natural formation of a grain boundary (GB) indicated by the dashed line in Fig.~\ref{Fig1}(a).
The GB with 24$^\circ$ misorientation angle acts as a Josephson barrier exhibiting a remarkably large critical current density  $j_0\sim 10^5\,{\rm A/cm}^2$ at 4.2\,K, typically.
Subsequently, a 70\,nm thick Au film is deposited in-situ by electron beam evaporation,  which provides resistive shunting to the Josephson junctions and protects the YBCO layer during patterning by focused ion beam (FIB) milling.
In this step, two bridges typically $w_{\rm J}\sim 300\,$nm wide and $l_{\rm J}\sim 300\,$nm long straddling the GB are formed to define the Josephson junctions intersecting the SQUID loop (see Fig.~\ref{Fig1}(b)).
For SQUID operation, a bias current $I_{\rm b}$ flows across the junctions (white arrows in Fig.~\ref{Fig1}(b)).
In addition, a typically $w_{\rm c}=100 - 200\,$nm wide and $l_{\rm c}= 200\,$nm long constriction is also patterned into the SQUID nanoloop; this provides the position with largest coupling for MNPs (see section \ref{subsec:calc-spin-sensitivity}).
Via a modulation current $I_{\rm mod}$ (black arrows in Fig.~\ref{Fig1}(b)) flowing through the constriction, the SQUID can be flux biased at its optimum working point, which also allows SQUID readout in flux locked loop (FLL) mode.\cite{Drung03}
The relevant geometric parameters for the SQUID loop are indicated in Fig.~\ref{Fig1}(c), and corresponding values for $l_{\rm J}$, $w_{\rm J}$, $l_{\rm c}$ and $w_{\rm c}$ for all five YBCO nanoSQUIDs are given in Table \ref{table1}.

Using FIB patterning, the lateral dimensions of the YBCO nanoSQUIDs can be controlled down to $\sim 50$ nm, providing a flexible and convenient way of tuning the size and geometry of the nanoloop.
This in turn determines its main parameters,\cite{Woelbing14} such as
(i) the maximum critical current $I_0$ of the Josephson junctions,
(ii) the total inductance $L$ with a geometric and kinetic contribution, the latter depending also on the film thickness, and (iii) the dimensions of the constriction, which determine the strength of maximum coupling of a MNP to the SQUID loop.\cite{Woelbing14}

\subsection{Measurement setup and high field operation}
\label{subsec:measurement}

Sensors are mounted in good thermal contact to the copper cold finger of a $^3$He refrigerator operative at 300\,mK $<T< $ 300\,K.
The refrigerator is introduced in a $^4$He cryostat hosting a vector magnet operating at a maximum sweeping rate of $\nu = 4.5\,$mT/s.
The vector magnet allows to carefully align the externally applied magnetic field $\bm H$ in the substrate (SQUID loop) plane and perpendicularly to the plane formed by the GB junctions (blue arrow in Fig.~\ref{Fig1}(b)).
In  this configuration, magnetic flux is coupled neither to the nanoSQUID loop nor to the Josephson junctions, allowing to operate the devices up to $\sim 1\,$T as demonstrated in Ref.~[\onlinecite{Schwarz13}].
To verify this, we have characterized a large number of bare nanoSQUIDs operating them in both open loop and FLL mode while sweeping $H$.
While the nanoSQUIDs are fully operative up to very large magnetic fields, we have observed the presence of abrupt changes in their response at $\mu_0 H \sim 1$\,T.
This behavior is still under investigation and is attributed to the entrance of Abrikosov vortices, probably stabilized at one or both sides of the constriction.
Measurements presented here have been obtained, however, at $\mu_0 H < 0.15\,$T where these effects play no role.

\subsection{Electrical characterization}
\label{subsec:characterization}

All devices presented here exhibited values of the maximum total critical current $I_{\rm c} = 2 I_0 \sim 500 - 600\,\mu$A at 4.2\,K, decreasing to $I_{\rm c} \sim 150 - 200\,\mu$A at 70\,K.
The response of the nanoSQUIDs at constant $I_{\rm b}$ can be modulated via $I_{\rm mod}$, allowing us to experimentally observe the $\Phi_0$-periodic response of the output voltage $V$ vs magnetic flux $\Phi$ in the SQUID loop ($\Phi_0$ is the magnetic flux quantum). 
From these measurements (see, e.g.,  Ref.\,[\onlinecite{Schwarz13}]) it is possible to determine the modulation currrent $I_{\rm mod,0}$ which is required to induce $1\,\Phi_0$. This yields the mutual inductance $M\equiv\Phi/I_{\rm mod}$ between the constriction and the SQUID.
The experimental determination of $M$ is paramount in order to quantify the flux $\Phi= V_{\rm out} M /R_{\rm f}$ coupled to the SQUID.
Here, $V_{\rm out}$ is the output voltage and $R_{\rm f}$ is the feedback resistance of the SQUID readout electronics operated in FLL mode ($R_{\rm f} = 3.3\,{\rm k}\Omega$, typically).

\begin{figure}[t]
\includegraphics[width=\columnwidth]{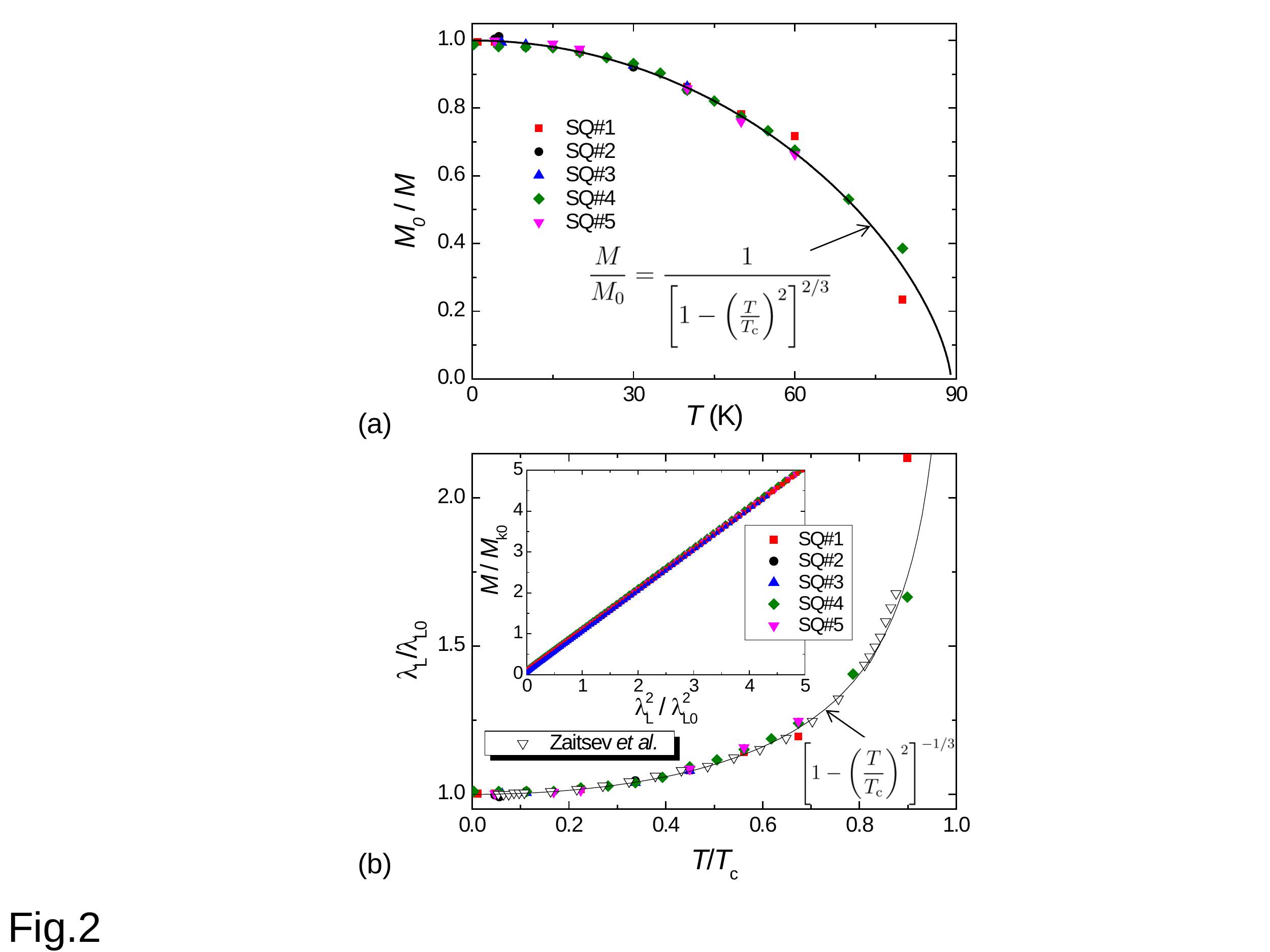}
\caption{
Temperature dependence of the mutual inductance $M$ and London penetration depth $\lambda_{\rm L}$ of all five YBCO nanoSQUIDs at $H=0$:
(a) Measured $T$ dependence of $1/M$ (symbols) normalized to $1/M_0$ (see Table \ref{table1}).
Solid line is a fit to the data with $T_c =$89\,K.
(b) $T$ dependence of $\lambda_{\rm L}$ extracted from data shown in (a) and numerical simulations of $M(\lambda_{\rm L}^2)$ as shown in the inset.
For comparison, the main graph includes data from Zaitsev {\it et al.} \cite{Zaitsev02}, with $\lambda_{\rm L0}=210\,$nm and $T_{\rm c}=92\,$K.}
\label{Fig2}
\end{figure}

\begin{table}[t]
\centering
\caption{Parameters for all five SQUIDs SQ$\#i$: geometric parameters of the SQUID layout ($w_{\rm J}$, $l_{\rm J}$, $w_{\rm c}$ and $l_{\rm c}$) and to $T=0$ extrapolated values for mutual inductances $M_0$,  $M_{\rm g}$, $M_{\rm k0}$ and London penetration depth $\lambda_{\rm L0}$.}
\label{table1}
\begingroup
\setlength{\tabcolsep}{2.0pt} 
\def\arraystretch{1.2}
\begin{tabular}{l | ccccccccccc}
& $w_{\rm J}$
& $l_{\rm J}$
& $w_{\rm c}$
& $l_{\rm c}$
& $M_0$
& $M_{\rm g}$
& $M_{\rm k0}$
& $\lambda_{\rm L0}$\\

&(nm)
&(nm)
&(nm)
&(nm)
&$\Phi_0$/mA
&$\Phi_0$/mA
&$\Phi_0$/mA
&(nm)  \\ \hline\hline

SQ$\#1$	& 380	& 330	& 500	& 220	& 0.29	& 0.02	& 0.27	& 243\\ 	
SQ$\#2$	& 270	& 255	&  80	& 265	& 1.17	& 0.08 	& 1.09	& 166\\ 	
SQ$\#3$	& 350	& 270	& 260	& 180 	& 0.58	& 0.02 	& 0.56	& 241\\ 	
SQ$\#4$	& 330	& 300	& 220  	& 250	& 0.44	& 0.04 	& 0.40	& 171\\ 	
SQ$\#5$	& 360	& 270	& 190  	& 190	& 0.48	& 0.03 	& 0.45	& 179\\ 	
\end{tabular}
\endgroup
\end{table}

%
Figure \ref{Fig2}(a) shows the measured $T$ dependence of $1/M$ for all five SQUIDs.
Here, $M$ is normalized to the extrapolated zero temperature value $M_0\equiv M(T=0)$ for each SQUID SQ$\#i$ (values for $M_{\rm 0}$ are listed in Table \ref{table1}).
The data are well approximated by $M/M_0=(1-t^2)^{-2/3}$, with the reduced temperature $t\equiv T/T_c$ and $T_c=89\,$K, which we will explain in the following.
Generally, $M$ contains both a geometric and a kinetic contibution, $M=M_{\rm g} + M_{\rm k}$.
$M_{\rm g}$ reflects the magnetic field produced by $I_{\rm mod}$, which is captured by the SQUID loop.
The kinetic part $M_{\rm k}$ reflects the contribution to the phase gradient of the superconductor wave function that is induced by the kinetic momentum of the Cooper pairs flowing along the constriction.
$M_{\rm k}$ is expected to be $T$-dependent through the Cooper pair density $n_{\rm s}(T)/n_{\rm s}(0)=\lambda_{\rm L0}^2/\lambda_{\rm L}^2(T)$,\cite{Tinkham96} with the London penetration depth $\lambda_{\rm L}$ and $\lambda_{\rm L0}\equiv\lambda_{\rm L}(T=0)$.
Hence, one expects $M_{\rm k} = M_{\rm k 0} \lambda_{\rm L}^2(T) / \lambda_{\rm L 0}^2$.
Generally, $M_{\rm g}$ can also be $T$-dependent, as the current density distribution across the constriction may vary with $\lambda_{\rm L}(T)$.
However, as the constriction width $w_{\rm c}$ is of the order of $\lambda_{\rm L0}$ (see Table \ref{table1} and determination of $\lambda_{\rm L0}$ values below), already for the lowest temperatures we can assume a rather homogeneous current density distribution across the constriction, which will not change with $T$.
In order to quantify the $T$-dependence of $M$ and its relation to $\lambda_{\rm L}(T)$, we performed numerical simulations of $M(\lambda_L)$ for the geometry of all five nanoSQUIDs, based on the London equations using the software package 3D-MLSI.\cite{Khapaev03}
The comparison of the measured values for $M(T)$ with the simulation results $M(\lambda_{\rm L})$ allows us to extract $\lambda_{\rm L}(T)$ and $\lambda_{\rm L0}$ for all five devices.
We note that the value of $\lambda_{\rm L0}$ extracted from the simulations crucially depends on the value for the constriction width $w_{\rm c}$.
Here, the largest uncertainty comes from the unknown value of the width of the damaged regions at the constriction edges due to FIB milling, which effectively reduces $w_{\rm c}$ by some value $\delta w_{\rm c}$.
This effect is strongest for SQ$\#2$ with the smallest $w_{\rm c}$.
Taking $\delta w_{\rm c}=20\,$nm, reduces the extracted value for $\lambda_{\rm L0}$ by 15\,\% for SQ$\#2$.
The inset of Fig.~\ref{Fig2} shows $M/M_{\rm k0}$ vs $(\lambda_{\rm L}/\lambda_{\rm L0})^2$.
Here, the data points for all five devices follow nicely the quadratic scaling with $\lambda_{\rm L}$, with almost invisible vertical shifts due to the small offset given by $M_{\rm g}/M_{\rm k0}$.
Values for $M_g$ and $M_{\rm k0}$ are listed in Table \ref{table1}.
We clearly see that $M_0=M_g+M_{\rm k0}$ is dominated by the kinetic contribution ($M_{\rm g} \sim(7\pm3)\,$\% of $M_{\rm k0}$).
$\lambda_{\rm L}(T)$ is displayed in Fig.~\ref{Fig2}(b), where we normalized $\lambda_{\rm L}$ to $\lambda_{\rm L0}$ and  $T$ to $T_{\rm c}=89\,K$.
The $T$-dependence of $\lambda_{\rm L}$ is roughly given by $\lambda_{\rm L}(T)=\lambda_{\rm L}(0)\,[1-t^2]^{1/3}$, leading to the observed $T$-dependence of $M$, since $M_{\rm g} \ll M_{\rm k}$.
Thus, due to the dominant kinetic contribution of $M$ in our devices, the measured $T$ dependence of $1/M$ (Fig.~\ref{Fig2}(a)) closely reflects the $T$ dependence of $n_{\rm s}$.
We note that the values for $\lambda_{\rm L0}$ vary from $\sim 170\,$nm to $\sim 240\,$nm for the five devices presented here.
Those values are significantly above the values $\lambda_{\rm L0}\sim 150\,$nm in the $a-b$ plane for YBCO single crystals.\cite{Buckel04}.
However, they are consistent with results by Zaitsev {\it et al.}\cite{Zaitsev02} obtained from microwave measurements of the absolute London penetration depth for epitaxially grown YBCO films and with results wich we obtained earlier for our YBCO nanoSQUIDs\cite{Schwarz13,Woelbing14} and thin films \cite{Thiel16}.
For comparison, we included in Fig.~\ref{Fig2}(b) results of one representative sample from Ref.~[\onlinecite{Zaitsev02}], which shows a $\lambda_{\rm L}(T)$ dependence that is very consistent with what we find for our devices presented here.

\begin{figure}[t]
\includegraphics[width=0.8\columnwidth]{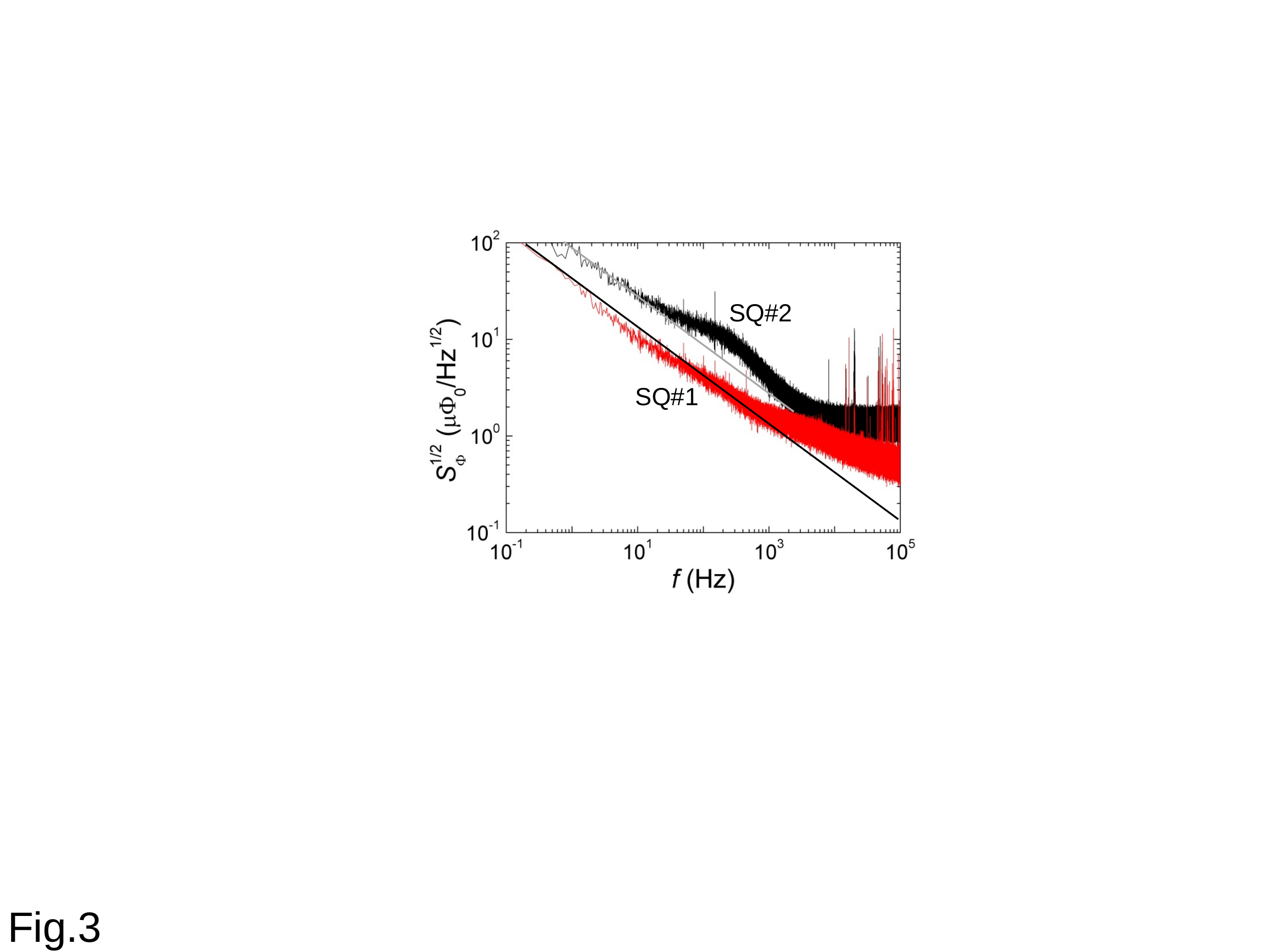}
\caption{
Rms spectral density of flux noise measured in FLL at 4.2\,K and $H=0$ for devices SQ$\#1$ and SQ$\#2$.
Solid lines indicate the 1/$f$ contributions.}
\label{Fig3}
\end{figure}

Finally, we have characterized the noise response of the devices in FLL mode obtaining very low values of the root-mean-square (rms) spectral density of flux noise $S^{1/2}_{\Phi}$.
Figure \ref{Fig3} shows data for SQ$\#$1 and SQ$\#$2 at $H=0$ and $T = 4.2$ K. 
The former exhibits the typical $S_\Phi\propto 1/f$ contribution, which dominates up to $\sim 1$ kHz where it starts to saturate reaching just $S^{1/2}_{\Phi} \sim 500$ n$\Phi_0$/Hz$^{1/2}$ at $100$\,kHz.
SQ$\#$2, on the other hand, exhibits also a $1/f$ contribution plus a broad peak at $\sim 200$\,Hz.
The noise in the white region is larger, in this case giving $S^{1/2}_{\Phi} \sim 1.2$ $\mu\Phi_0$/Hz$^{1/2}$ at $100$ kHz.
Both the presence of peaks in the noise spectra and excess $1/f$ contributions are typically found in these devices.\cite{Koelle99,Schwarz15}
These effects have been attributed to $I_0$ fluctuations in the GB junctions and to the existence of ubiquitous magnetic fluctuators either at the STO/YBCO interface or in the GB junctions.\cite{Schwarz15}

Operation in external magnetc fields and at variable temperature has been investigated experimentally by measuring the noise of SQ$\#$1 at $-100\,{\rm mT} <\mu_0H < 400\,$mT and $0.3\,{\rm K}< T< 50\,$K.
Similarly to the spectra shown in Fig. \ref{Fig3}, the flux noise is dominated by a large $1/f$ contribution exhibiting the presence of peaks at frequencies that depend on both $T$ and $H$.
Although no systematic $T$- or $H$-dependence has been found,\cite{Schwarz15} we can state that noise spectra are only weakly affected by the application of external magnetic fields or by the operation at higher temperatures.
As a matter of fact, $S^{1/2}_{\Phi}$ values at $100$\,kHz do not change by more than a factor $4 - 5$.

\subsection{Calculation of coupling and spin sensitivity}
\label{subsec:calc-spin-sensitivity}

\begin{figure}[b]
\includegraphics[width=\columnwidth]{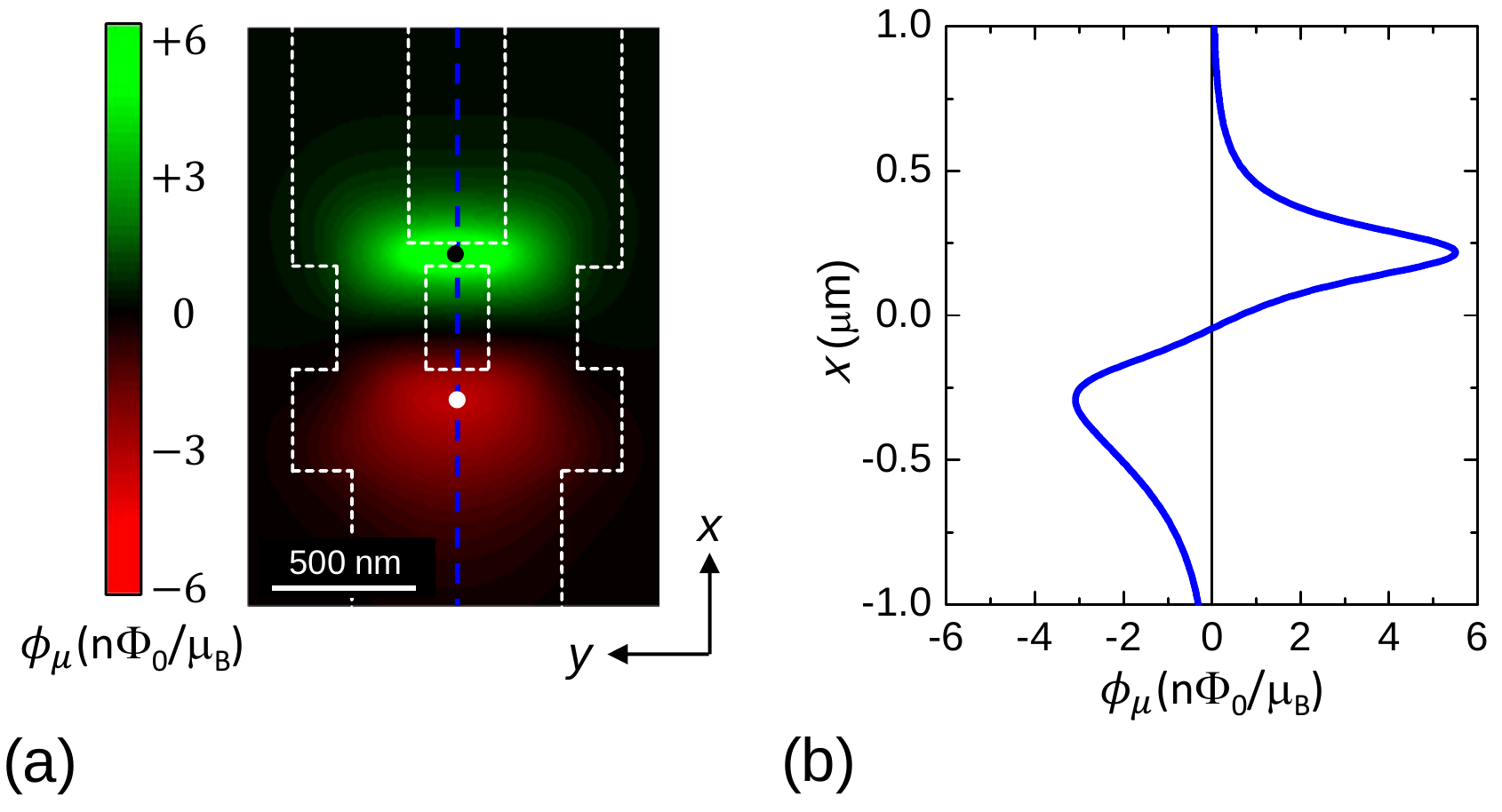}
\caption{
(a) $\phi_{\mu}(x,y)$ contour plot of a typical nanoSQUID, 10\,nm above the surface of the 70\,nm thick Au layer, for $\hat{\bm e}_\mu\, || \,\hat{\bm e}_x$. 
White dashed lines indicate the contour of the chosen nanoSQUID geometry with an 80\,nm wide constriction.
Black and white dots indicate the positions of maximum coupling, i.e., $\phi_{\mu} =5.5$  and $-3.1\,{\rm n}\Phi_0/\mu_{\rm B}$, respectively.
The blue dashed line indicates the position of the line scan shown in (b).
(b) $\phi_\mu(x)$ along the dashed line in (a). Note that the $x$-axis (vertical axis) in (a) and (b) coincide.
}
\label{Fig4}
\end{figure}

In order to estimate the regions of maximum coupling for MNPs above the surface of the sensors, we perform numerical simulations based on the London equations to calculate the coupling factor $\phi_\mu({\bm r},\hat{\bm e}_\mu)$.
This quantity expresses the amount of magnetic flux coupled into the nanoSQUID loop  per magnetic moment of a point-like MNP with its magnetic moment $\bm\mu=\mu\hat{\bm e}_\mu$ oriented along  $\hat{\bm e}_\mu$ and located at position $\bm r$.
$\phi_\mu$ was calculated using 3D-MLSI\cite{Khapaev03} to obtain the magnetic field ${\bm B}_J({\bm r})$ at position $\bm r$ induced by a current $J$ circulating in a 2-dimensional sheet around the SQUID hole, taking into account the lateral geometry of the SQUID.
As shown in Refs.~[\onlinecite{Nagel11}] and [\onlinecite{Woelbing14}], $\phi_\mu(\bm r,\hat{\bm e}_\mu)$ can then be obtained via
%
\begin{equation}
\phi_\mu(\bm r,\hat{\bm e}_\mu) = - \hat{\bm e}_\mu \cdot \bm B_J(\bm r) / J\;.
\label{eq:phi_mu}
\end{equation}
%
This calculation was done for 11 current sheets spread equally across the film thickness as described in Ref.~[\onlinecite{Woelbing14}].
The resulting coupling factors were averaged for each position $\bm r$, resulting in position-resolved maps $\phi_\mu(x,y)$ in the plane parallel to the SQUID loop plane, as shown in Fig.~\ref{Fig4}(a).
Figure \ref{Fig4}(b) shows a line-scan $\phi_{\mu}(x)$ calculated along the dashed line in Fig.~\ref{Fig4}(a).
Results plotted in Fig.~\ref{Fig4} have been calculated for a vertical distance $z=80\,$nm above the YBCO surface, i.e.,  10\,nm above the Au surface, for a device with geometry similar to SQ$\#$2.
In these calculations we have assumed $\hat{\bm e}_\mu$ parallel to the externally applied magnetic field $H$ (along the $x$ direction).
Note that $\phi_\mu$ reverses it sign upon going from the constriction to the opposite side of the SQUID loop.
This is simply related to the direction of the flux lines coupled to the nanoloop and makes no difference for the measurements performed here.
As $\lambda_{\rm L0}$ varies from $\sim 170 - 250\,$nm for our devices, we have performed simulations of $\phi_\mu$ for variable $\lambda_{\rm L0}$.
In contrast to the scaling of $M(\lambda_{\rm L})$, we find only a very weak dependence of $\phi_\mu(\lambda_{\rm L})$.
Hence, for the calculations of $\phi_\mu$ presented below, we fixed $\lambda_{\rm L}$ to 250\,nm to be consistent with our earlier  work \cite{Woelbing14}.

Regions of maximum $|\phi_\mu|$ are found at the constriction ($\phi_\mu = 5.5\,{\rm n}\Phi_0/\mu_{\rm B}$, at the position of the black dot in Fig.~\ref{Fig4}(a)) and at the opposite side of the nanoSQUID loop ($\phi_\mu = -3.1\,{\rm n}\Phi_0/\mu_{\rm B}$, at the position of the white dot  in Fig.~\ref{Fig4}(a)); $\mu_{\rm B}$ is the Bohr magneton.
A particle located at the constriction is better coupled as this is the region with smallest linewidth of the SQUID.
Accordingly, more flux lines can be captured through the nanoloop.

Experimental values of $S^{1/2}_{\Phi}$ and the calculated  $\phi_\mu$ allow estimating the expected spin sensitivity.
This is the figure of merit of nanoSQUID sensors, defined as
%
\begin{equation}
S^{1/2}_{\mu} =  S^{1/2}_{\Phi}  /|\phi_\mu|.
\label{eq:S_mu}
\end{equation}
%
For a point-like particle on top of the constriction of SQ$\#$2 at $z=80\,$nm, we obtain $S^{1/2}_{\mu} \sim 220\,\mu_{\rm B}/{\rm Hz}^{1/2}$ at 100\,kHz.
This means that $220\,\mu_{\rm B}$ fluctuating at 100\,kHz can be detected in a 1\,Hz bandwidth.

\section{Co MNP growth}
\label{sec:MNP-patterning}

Polycrystalline cobalt MNPs have been grown by FEBID in a dual-beam system from FEI (models Helios 600 and 650).
The focused electron beam is used to take SEM images of the nanoSQUID and spot the precise location where the Co MNP is desired to be grown.
The precursor gas Co$_2$(CO)$_8$  is supplied locally with a gas injection system that approaches a needle to a distance $\sim 150\,\mu$m from the site of interest.
The base pressure of the chamber is $2\times 10^{-6}\,$mbar, and increases to $\sim 1.5\times 10^{-5}\,$mbar when the precursor valve is open.
The electron beam is scanned on the selected area using a small current (25\,pA) to ensure a good spatial resolution, and with low voltage (5\,kV) to produce a material with moderate purity $\sim 60\,{\rm at.}\%$; a higher purity of $90\,\%$ is possible, but then the sample is very prone to oxidation in ambient conditions.\cite{Pablo-Navarro16}

\begin{figure}[t]
\includegraphics[width=0.75\columnwidth]{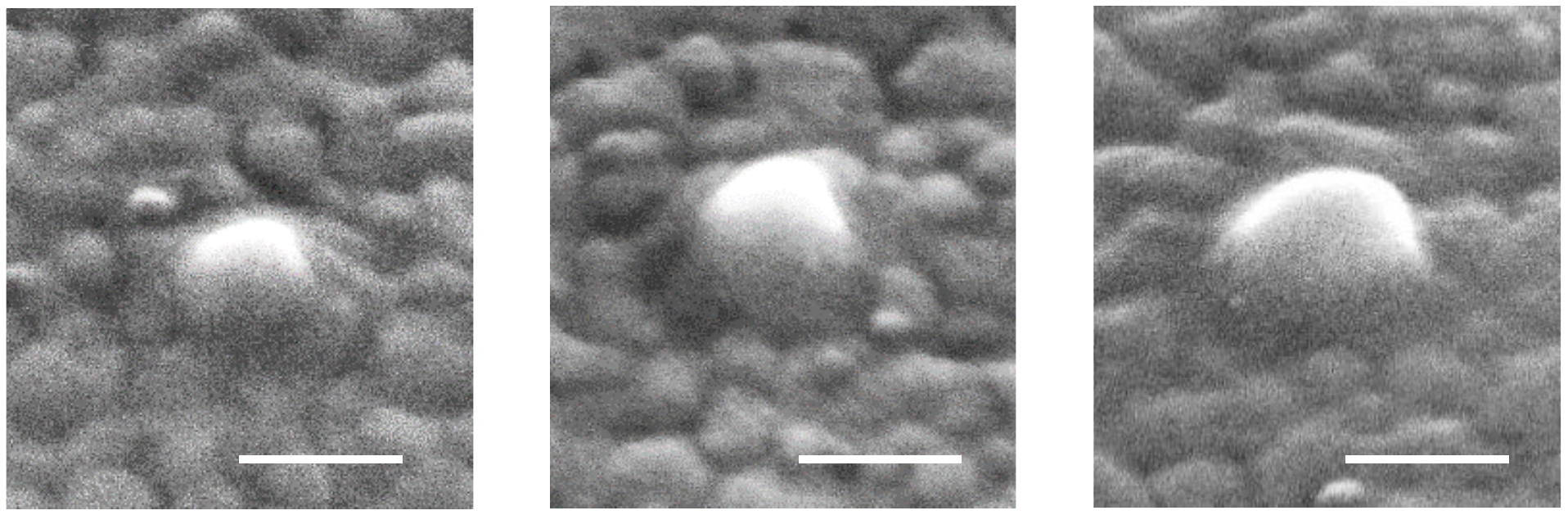}
\caption{SEM images of three typical cobalt nanoparticles deposited by FEBID on top of a YBCO/Au bilayer. Scale bar is $100\,$nm.
}
\label{Fig5}
\end{figure}

Three FEBID-Co nanoparticles grown on top of a YBCO/Au bilayer are displayed in Fig.~\ref{Fig5}, showing a high degree of control over the geometrical volume $V_{\rm geo}$ of the particles.
These particles have been obtained by scanning the electron beam on a $10\,$nm diameter circle leading to the formation of a spherical cap-like MNP with geometrical diameter $d_{\rm geo}$ and thickness (height) $t_{\rm geo}$. 
From left to right in Fig.~\ref{Fig5}, these MNPs have $d_{\rm geo}/t_{\rm geo}$ $\sim 65/30$, $\sim 85/40$ and $\sim 115/60\,$\,nm/nm.
Still, the likely presence of a magnetically dead/paramagnetic layer does not allow the precise determination of the real magnetic volume $V_{\rm mag}$ of the Co particles from SEM images.
The dead layer might arise at the first stage of the growth process due to a likely lower concentration of Co.\cite{Fernandez-Pacheco09}
Partial oxidation at the surface of the particle might also lead to a thin antiferromagnetic CoO$_x$ layer.\cite{Pablo-Navarro16}
In Sec.~\ref{subsec:spin-sensitivity-coupling} $V_{\rm mag}$ will be estimated by combining the calculated coupling between the MNP and the SQUID nanoloop and the experimentally measured magnetic flux.

\begin{figure}[t]
\includegraphics[width=\columnwidth]{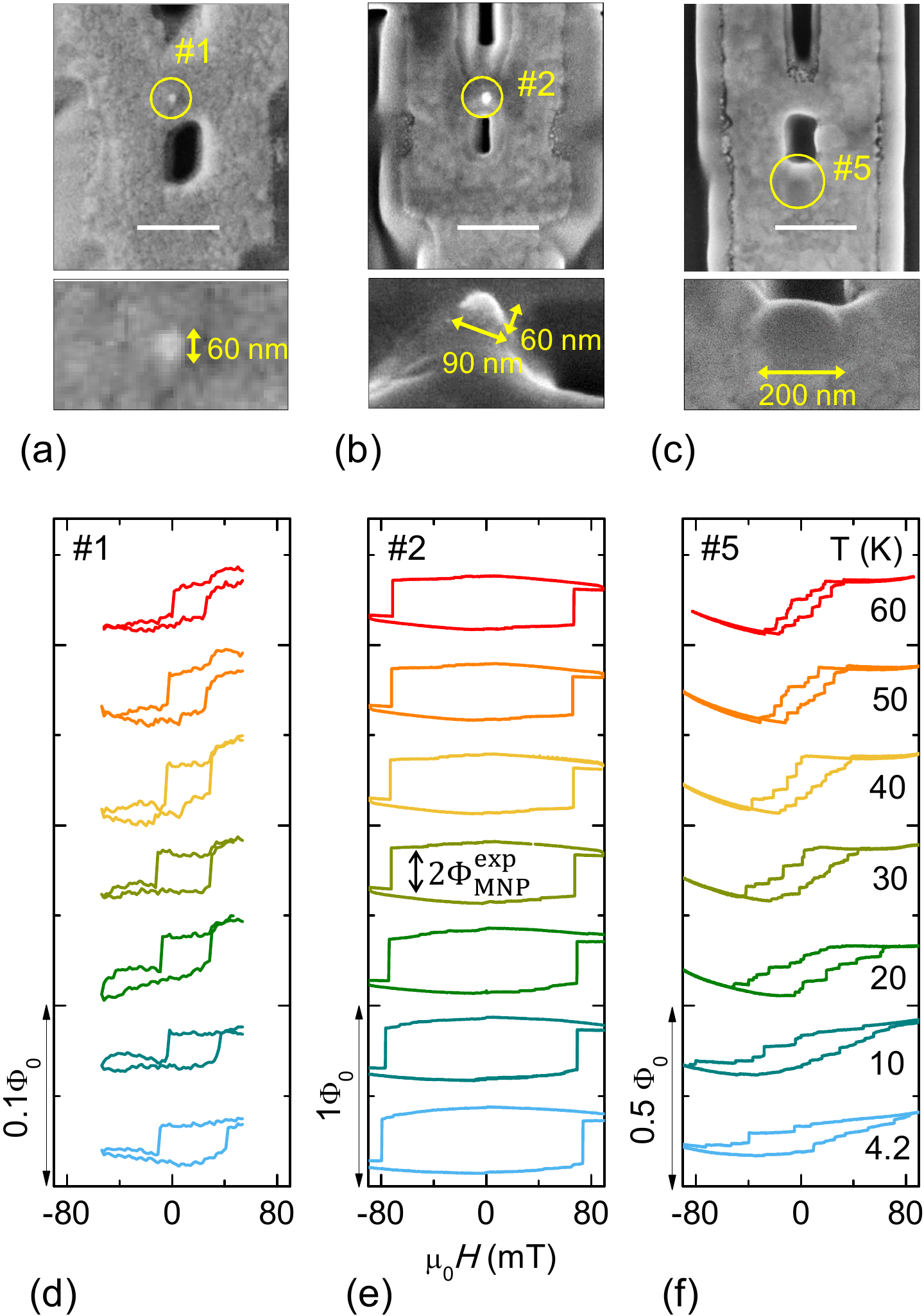}
\caption{(a)$-$(c) SEM images of devices SQ$\#1$, SQ$\#2$ and SQ$\#5$, respectively (upper panels); scale bars correspond to 500\,nm. 
Co MNPs are highlighted by circles and shown in zoomed view in the bottom panels (tilted  images in (b) and (c)).
(d)$-$(f) Representative hysteresis curves $\Phi(H)$ measured for MNP $\#1$, $\#2$ and $\#5$, respectively, at different temperatures as indicated in (f). The field sweep rate was $\nu = 4.5\,$mT/s in (d) and $\nu = 0.45\,$mT/s in (e) and (f).
Curves are vertically shifted for clarity.
The vertical axes are in units of magnetic flux coupled to the nanoSQUID; notice the different scales.}
\label{Fig6}
\end{figure}

FEBID-Co nanoparticles have been grown as described above at the precise positions where $\phi_\mu$ is maximum.
SEM images of three representative samples, SQ$\#1$, SQ$\#2$ and SQ$\#5$,  are shown in Fig.~\ref{Fig6}(a), (b) and (c), respectively.

Particles $\#1$, $\#2$ and $\#3$ are similar to those shown in Fig.~\ref{Fig5}.
Their estimated geometrical dimensions correspond to $d_{\rm geo}/t_{\rm geo} \sim 60/40$, $\sim 90/60$ and $\sim 50/35$\,nm/nm, respectively.
Their geometrical volume is then obtained as that of a spherical cap, i.e., $V_{\rm geo} = \frac{\pi}{6} t_{\rm geo}(\frac{3}{4}d^2_{\rm geo} +t^2_{\rm geo})$.
Particle $\#1$ is placed above the 500\,nm-wide constriction of SQ$\#1$ (Fig.~\ref{Fig6}(a)), whereas particle $\#2$ lies on top of the much narrower 80\,nm-wide constriction of SQ$\#2$ (Fig.~\ref{Fig6}(b)), and particle $\#3$ sits on SQ$\#3$ with intermediate constriction width $w_{\rm c}=260\,$nm.
This entails clear differences between their respective coupling factors, being largest for particle $\#2$, as will be discussed in Sec.~\ref{subsec:spin-sensitivity-coupling}.

Particles $\#4$ and $\#5$ are, on the other hand, disc-shaped.
They were grown by scanning the electron beam on $100$ and $200\,$nm diameter circles, respectively.
Their geometrical thickness is $t_{\rm geo} \sim 35$ nm, as determined by atomic force microscopy performed directly on the surface of the sensors.
From these measurements we also conclude that the surface of the MNPs is very smooth (5\,nm  roughness).
In this case, $V_{\rm geo}= \frac{\pi}{4} t_{\rm geo}d^2_{\rm geo}$ has been calculated as that of a cylinder.

We highlight that the larger discs $\#4$ and $\#5$, have been deposited  close to the edge of the nanoloop opposite to the constriction.
The reason is that this region provides a smoother Au surface, less affected by FIB milling effects at the edges.
As shown in Fig.~\ref{Fig4}, this region still offers large values of $\phi_\mu$.
Together with their larger volumes, this provides reasonable magnetic signals as we will see in the following.

\section{Magnetization measurements}
\label{sec:magnetization-measurements}

Magnetization hysteresis loops of Co MNPs, i.e. change of magnetic flux $\Phi$ coupled to the nanoSQUID vs applied magnetic field $H$, of the different samples have been obtained by sweeping $H$ at different temperatures while operating the nanoSQUIDs in FLL mode.
Except for the measurements presented in Sec.~\ref{subsec:angular-dependence}, $H$ was always applied perpendicular to the GB plane.
Some representative measurements performed with SQ$\#1$, SQ$\#2$ and SQ$\#5$,  are shown in Fig.~\ref{Fig6}(d), (e) and (f) respectively.
These curves have been obtained at the same temperatures as indicated in Fig.~\ref{Fig6}(f).
All particles exhibit hysteretic behavior.
The magnetic signal of each particle saturates at different values of the external magnetic field in the range $40\,{\rm mT} <\mu_0 H < 80\,$mT.
When sweeping back the magnetic field from the fully saturated state, abrupt steps indicate the onset of an irreversible process of magnetization reversal.
In all cases, the observed switching fields depend on temperature, suggesting the occurrence of a thermally activated magnetization reversal process.

However, clear differences are observed in measurements on different samples.
Hysteresis curves corresponding to $\#1$, $\#2$ and $\#3$ are square shaped, suggesting that particles remain in the (quasi) single-domain state while $H$ is swept.
This does not necessarily mean that particles are uniformly magnetized. Non-uniformities are likely to appear at the edges so to reduce the total magnetostatic energy.\cite{Cowburn00,Han07}

In contrast to this, measurements obtained with $\#4$ and $\#5$ exhibit a number of reproducible steps, suggesting that magnetization reversal is assisted by the formation of more complicated multi-domain magnetic states. 
Owing to the circular shape of the particles, flux-closure magnetic states such as vortices might be stabilized at equilibrium or nucleate when sweeping the magnetic field.\cite{Guslienko08,Fruchart05}
These measurements will be analyzed in more detail elsewhere.

\subsection{Magnetic flux signals}
\label{subsec:spin-sensitivity-coupling}

\begin{table*}[t]
\centering
\caption{Experimentally measured $\Phi_{\rm MNP}^{\rm exp}$ (with an rms noise amplitude $\sim 1{\rm m}\Phi_0$), geometric MNP parameters $d_{\rm geo}$, $t_{\rm geo}$ and $V_{\rm geo}$ (determined from SEM images with an estimated error $\pm 10\,$nm in $d_{\rm geo}$ and $t_{\rm geo}$), calculated values of $|\phi_{\rm MNP}|$ (for $\Phi_{\rm MNP}^{\rm theo}=\Phi_{\rm MNP}^{\rm exp}$), magnetic MNP parameters $V_{\rm mag}$ and $t_{\rm mag}$ (determined from $\Phi_{\rm MNP}^{\rm exp}$ and $\phi_{\rm MNP}$), and estimated magnetic moment $\mu_{\rm MNP}$ for each particle.}
\label{table2}
\begingroup
\setlength{\tabcolsep}{7.0pt} 
\def\arraystretch{1.5}
\begin{tabular}{l | cccccccc}
& $\Phi_{\rm MNP}^{\rm exp}$
& $d_{\rm geo}$
& $t_{\rm geo}$
& $V_{\rm geo}$
& $|\phi_{\rm MNP}|$
& $V_{\rm mag}$
& $t_{\rm mag}$
& $\mu_{\rm MNP}$ \\

&(m$\Phi_0$)
&(nm)
&(nm)
&($\times 10^{-16}$cm$^3$)
&(n$\Phi_0/\mu_{\rm B}$)
&($\times 10^{-16}$cm$^3$)
&(nm)
&($\times 10^{6}\mu_{\rm B}$) \\ \hline\hline

$\#1$	& 10	 		&60		&40	&$0.9\pm0.2$	&$3.0\pm0.2$&$0.4\pm0.1$&$21\pm6$	&$3.3\pm0.9$\\
$\#2$	& 110		&90 		&60	&$3.0\pm0.5$	&$4.9\pm0.6$&$2.4\pm0.5$&$52\pm8$	&$23\pm6$ \\
$\#3$	& 5.5		&50 		&35	&$0.6\pm0.2$	&$3.9\pm0.3$&$0.15\pm0.04$&$14\pm5$&$1.4\pm0.4$	 \\
$\#4$	& 24	  		&100	&35	&$2.7\pm0.5$	&$3.0\pm0.2$&$0.9\pm0.2$&$11\pm3$		&$8.0\pm2.0$\\
$\#5$	& 80	  		&200	&35	&$11\pm2$	&$2.7\pm0.2$&$3.2\pm0.6$&$10\pm2$	&$30\pm7$ \\
\end{tabular}
\endgroup
\end{table*}

The maximum experimentally detected magnetic flux $\pm\Phi_{\rm MNP}^{\rm exp}$ coupled by a fully saturated Co MNP to the nanoSQUID depends on the position, size and saturation magnetization $M_s$ of the MNP and on the specific geometry of the nanoSQUID through the magnetic coupling.
This can be appreciated in Fig.~\ref{Fig6}(d), (e) and (f) by observing the differences in $\Phi_{\rm MNP}^{\rm exp}$ for different samples or in Table \ref{table2}, where the values of $\Phi_{\rm MNP}^{\rm exp}$ are summarized.
$\Phi_{\rm MNP}^{\rm exp}$ can be compared with the expected signal calculated as $\Phi_{\rm MNP}^{\rm theo} =|\phi_{\rm MNP}| V M_s$.
Here, $V$ is the volume of the particle, $M_s = p M_s^{\rm Co}$ where  $p=(60\pm 10)\,{\rm at.}\%$ is the expected concentration of Co atoms, and $M_s^{\rm Co} = 1.4 \times 10^6 $ A/m is the saturation magnetization of cobalt.\cite{Grimsditch97}
$\phi_{\rm MNP}$ is the averaged coupling factor for each nanoSQUID across the particle volume, given by
%
\begin{equation}
\phi_{\rm MNP}= \frac{\int_{V} \phi_\mu ({\bf r}) dV }{V}\;.
\label{eq:Phi_MNP}
\end{equation}
%
In all cases, taking  $V=V_{\rm geo}$ in the above formula yields values of $\Phi_{\rm  MNP}^{\rm theo}$ larger than the experimental ones.
This fact suggests an effective magnetic volume $V_{\rm mag}$ smaller than $V_{\rm geo}$ estimated from the SEM images and the AFM measurements.
$V_{\rm mag}<V_{\rm geo}$ is reasonable considering that the FEBID process might lead to an effectively dead magnetic layer as discussed in section \ref{sec:MNP-patterning}.
We have estimated $V_{\rm mag}$ as the volume required in order to obtain $\Phi_{\rm MNP}^{\rm theo}=\Phi_{\rm MNP}^{\rm exp}$.
The magnetic thickness $t_{\rm mag}$ is then calculated by assuming $d_{\rm mag} = d_{\rm geo}$.
Estimated values are given in Table \ref{table2} together with the measured $d_{\rm geo}$, $t_{\rm geo}$ and $V_{\rm geo}$.
The table also provides the total resulting estimated magnetic moment  per particle $\mu_{\rm MNP} = V_{\rm mag} M_s$ and the values of $\phi_{\rm MNP}$ calculated by averaging across $V = V_{\rm mag}$.

In case of particle $\#2$ a volume averaged magnetic coupling of $\phi_{\rm MNP}=4.9$ n$\Phi_0/\mu_{\rm B}$ has been calculated.
This value results when integrating over the magnetic volume of  MNP $\#2$ assuming it lies on top of the Au layer, i.e., at $z=70$ nm.
We note that due to FIB-induced rounding of the nanoSQUID patterned edges, the Au thickness and hence $z$ may vary across the constriction.
This effect becomes especially important in nanoSQUID $\#2$ with the smallest constriction width ($w_c \sim 80$ nm) where the particle is deposited very close to the edge (see bottom panel of Fig.~\ref{Fig6}(b)).
In this case, assuming a reasonable value of, e.g., $z=35$ nm would yield $\phi_{\rm MNP}=7.3$ n$\Phi_0/\mu_{\rm B}$ obtained upon integration over $V_{\rm mag}= (1.6 \pm 0.3) \times 10^{-16}$ cm$^3$.
This translates into an estimated magnetic thickness of $t_{\rm mag} = 40 \pm 7$ and $\mu_{\rm MNP} = (15 \pm 4) \times 10^{6}$ $\mu_{\rm B}$.

As it can be seen, the effective magnetic volume of each particle is smaller than the geometrical one by a factor $\sim 3$ on average.
Put another way, the dead magnetic layer amounts to  $20 - 25$ nm roughly.
Alternatively, $\Phi_{\rm MNP}^{\rm theo}$ and $\Phi_{\rm MNP}^{\rm exp}$ could agree if we assume that these particles have a much lower amount of Co atoms.
In this case, the Co purity can be estimated by assuming $V = V_{\rm geo}$, leading to just  $p \sim 20\,{\rm at.}\%$.
We consider this latter scenario as unrealistic, as such a low Co concentration would yield a purely paramagnetic material.\cite{Gabureac10,DeTeresa16}

We note the  high signal-to-noise ratio of our hysteresis loop measurements (the rms noise amplitude of $\Phi_{\rm MNP}^{\rm exp}$ amounts to $\lapprox$ $1{\rm m}\Phi_0$).
This  is due to the high spatial resolution achieved with FEBID growth of the MNPs directly on top of the FIB-patterned constrictions in the nanoSQUIDs.
For SQ$\#2$ with only 80\,nm constriction width, the largest value of the averaged coupling factor is achieved.
In this case, MNPs having a total magnetic moment of just $\sim 10^5\,\mu_{\rm B}$ would still provide a measurable signal.

\subsection{Temperature dependence}
\label{subsec:T-dependence}

\begin{figure}[b]
\includegraphics[width=0.6\columnwidth]{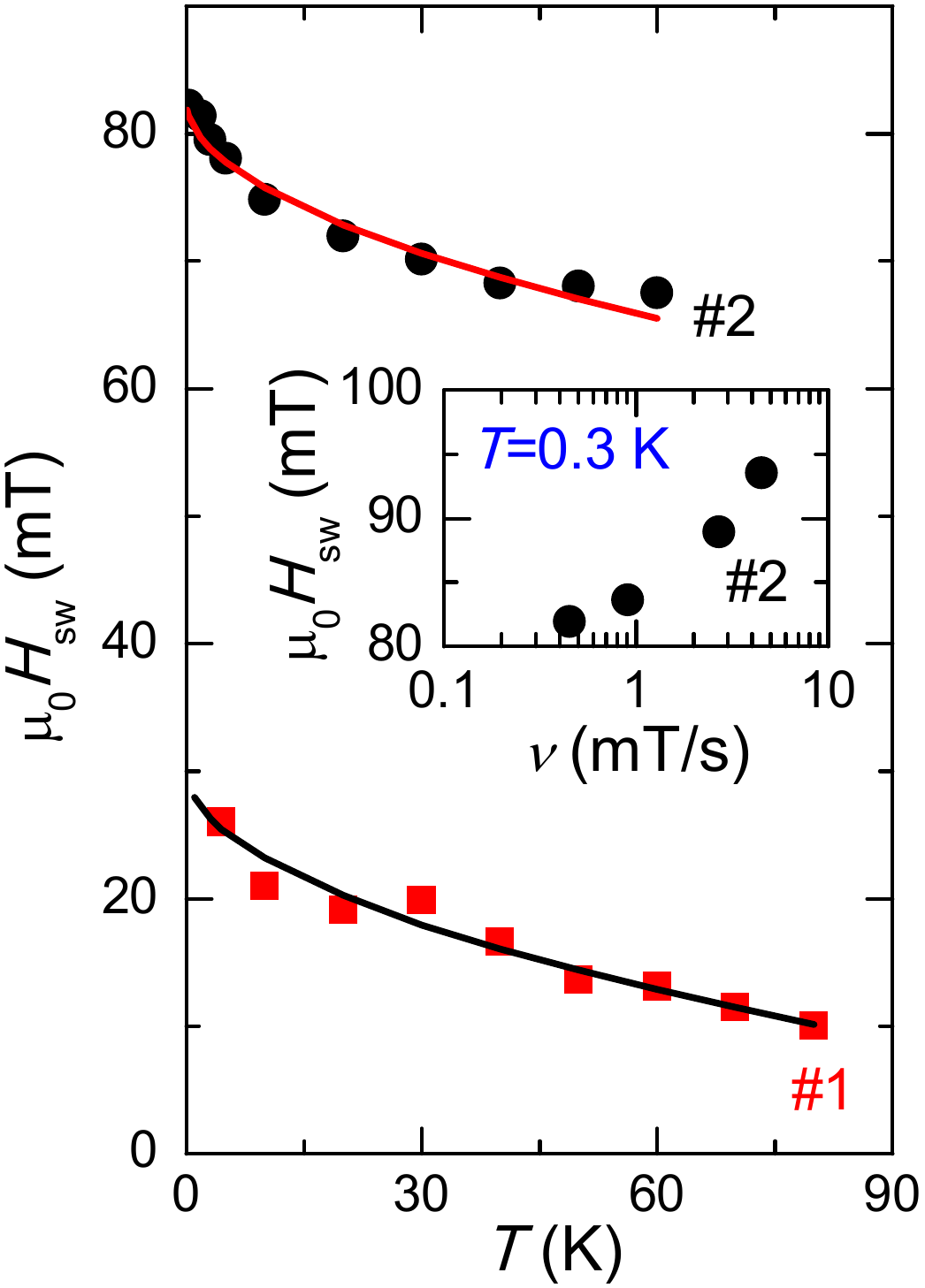}
\caption{Temperature dependence of $H_{\rm sw}$ for particle $\#1$ ($\nu = 4.5\,$mT/s) and $\#2$ ($\nu = 0.45\,$mT/s).
Solid lines are fits to the Kurkij\"arvi model, Eq.~(\ref{eq:kurkijarvi}), for a thermally activated process over an energy barrier.
The inset shows $H_{\rm sw}$ vs field sweeping rate at $0.3$ K for particle $\#2$.}
\label{Fig7}
\end{figure}

The $T$ dependence of the switching magnetic field $H_{\rm sw}$ is analyzed in the following.
$H_{\rm sw}$ is defined as $H_{\rm sw}=(H^+_{\rm sw}+H^-_{\rm sw})/2$ where $H^{+(-)}_{\rm sw}$ is that at which irreversible jumps are observed in the hysteresis curves when sweeping up (down) the field.
Except for these jumps occurring at $H^+_{\rm sw}$ and $H^-_{\rm sw}$, the nanoSQUID output signal vs $H$ is reversible.
%
$H_{\rm sw} (T)$ values for particles $\#1$ and $\#2$ are plotted in Fig.~\ref{Fig7} showing that $H_{\rm sw}$ decreases with increasing $T$.
As mentioned above, this behavior is typical for a single-domain particle if its magnetization reversal is assisted by thermal fluctuations.
Such fluctuations allow the magnetization to overcome the energy barrier $U_0$ created by the magnetic anisotropy.
Being an stochastic process, $H_{\rm sw}$ should depend on both the temperature $T$ and the field sweeping rate $\nu$.
This is further confirmed by the fact that $H_{\rm sw}$ increases with increasing  $\nu$, as shown in the inset of Fig.~\ref{Fig7} where data were taken at $T=0.3$ K.
Within the N\'eel-Brown model of magnetization reversal,\cite{Neel49,Brown63} the mean switching field can be obtained from the model of Kurkij\"arvi\cite{Kurkijarvi72,Garg95,Gunther94}
%
\begin{equation}
\mu_0 H_{\rm sw} = \mu_0 H_{\rm sw}^0 \left\{   1 - \left[  \frac{k_{\rm B} T}{ U_0}  \ln \left( \frac{c T}{  \nu}  \right)  \right]^{1/\alpha}  \right\}\;,
\label{eq:kurkijarvi}
\end{equation}
%
where $c= H_{\rm sw}^0 k_{\rm B}  / \tau_0 \alpha U_0 \varepsilon^{\alpha-1}$.
$H_{\rm sw}^0$ is the switching field at $T=0$, $\varepsilon = 1- H_{\rm sw}/H_{\rm sw}^0$, $\tau_0$ is an attempt time, $k_{\rm B}$ is the Boltzmann constant and $\alpha$ varies usually between $1 - 2$.\cite{Fruchart05}
Experimental data are fitted by Eq.~(\ref{eq:kurkijarvi}) as shown by the solid lines in Fig.~\ref{Fig7} where best fits are found for $\alpha = 2$.
For both particles, $\tau_0=10^{-10}$\,s has been used, although it influences only marginally the fits.
We found $U_0/k_{\rm B} = 3.8\times 10^3\,$K and $\mu_0 H_{\rm sw}^0 = 30\,$mT for particle $\#1$ and  $U_0/k_{\rm B} = 3.2\times 10^4\,$K and $\mu_0 H_{\rm sw}^0 = 83\,$mT for particle $\#2$.
This is of the same order of magnitude as  $U_0/k_{\rm B} = 6.8 \times 10^3$\,K and $2.7 \times 10^4 $\,K, obtained by Wernsdorfer {\it et al.} \cite{Wernsdorfer95b} for elliptical polycrystalline cobalt particles with dimensions $80 \times 50 \times 30$ and $150 \times 80 \times 30$ nm$^3$, respectively.

The energy barrier can be translated into a phenomenological activation volume $V_{\rm act} = U_0/ \mu_0 H_{\rm sw}^0 M_s$.
Calculated values for particle $\#1$ and $\#2$ yield $V_{\rm act} \sim$ 2 and 6 $\times 10^{-18}\,{\rm cm}^3$, respectively.
These values are just $\sim 6\,\%$ and $3\,\%$ of $V_{ \rm mag}$ for each particle, respectively, suggesting that magnetization reversal is triggered by a nucleation process followed by propagation of domain walls.
According to this picture, magnetization reversal initiates within a small region of volume $\sim V_{\rm act}$. This is followed by a rapid (ps - ns) propagation of the reversed magnetization through the whole volume of the particle.
This process cannot be distinguished from pure coherent magnetization reversal by only inspecting the hysteresis curves as both mechanisms, i.e., nucleation and propagation, take place within the experimental field step-size.

\begin{figure}[t]
\includegraphics[width=7cm]{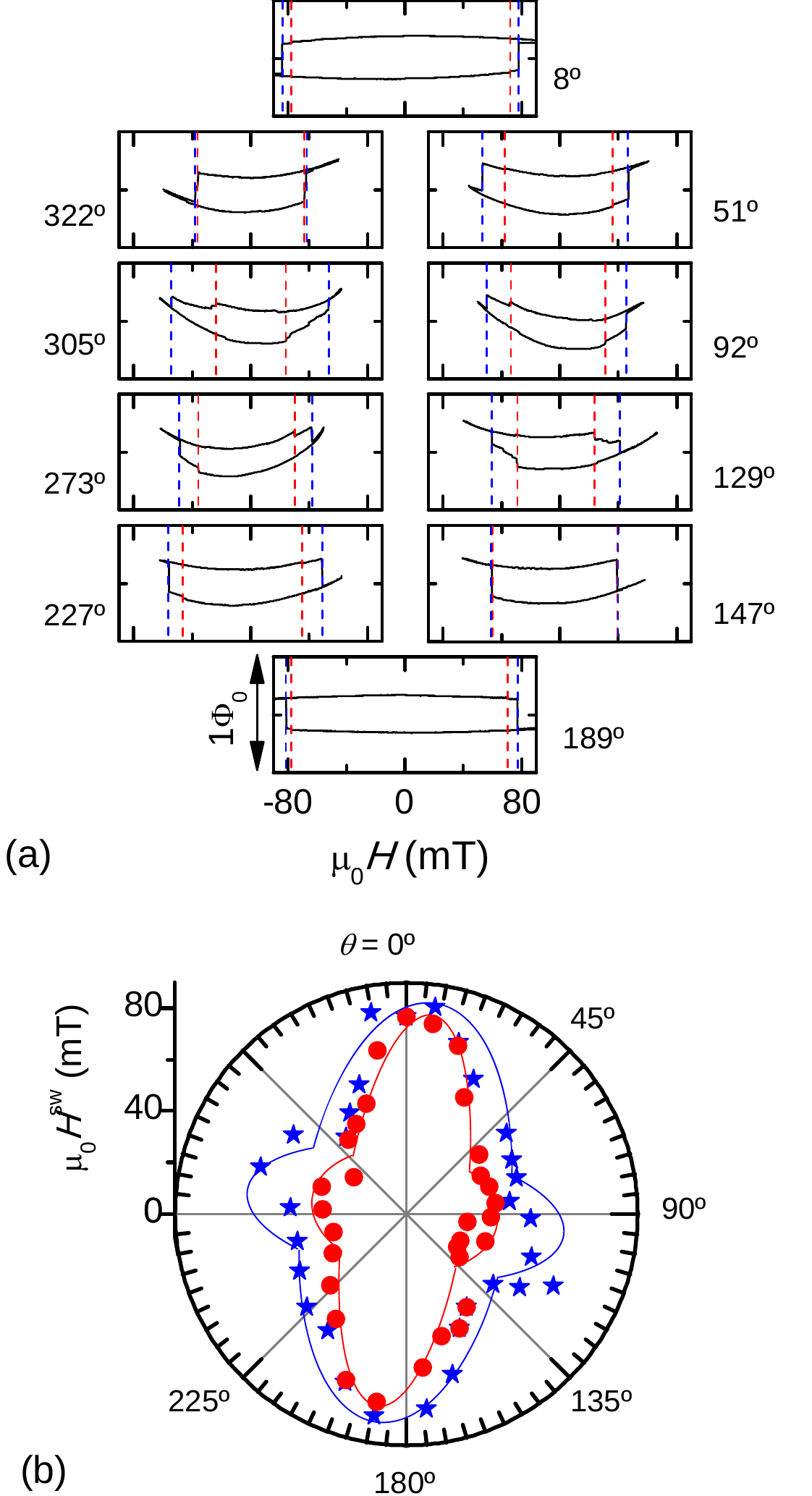}
\caption{Angular dependence of magnetization reversal for MNP $\#2$ obtained by rotating $H$ within the substrate plane at $T=4.2$\,K.
(a) $\Phi(H)$ Hysteresis curves obtained at different values of $\theta$ indicated at each panel.
Red (blue) lines indicate the position of the first (last) switching step.
The height of each graph corresponds to  $1\,\Phi_0$.
(b) $H_{\rm sw}(\theta)$ polar plot. For each value of $\theta$, the magnitude of the corresponding switching field is represented by the distance from dots (stars) to the origin. Dots: first step; stars: last step.
The red and blue lines are guides to the eye highlighting the fourfold symmetry.}
\label{Fig8}
\end{figure}

\subsection{Angular dependence}
\label{subsec:angular-dependence}

In order to gain a deeper insight into the mechanisms leading to magnetization reversal,  we performed magnetization measurements by rotating the externally applied magnetic field by an angle $\theta$ in the plane of the nanoSQUID loop (substrate plane).
Results are shown in Fig.~\ref{Fig8}(a) where few representative $\Phi(H)$ hysteresis curves are shown for different values of $\theta$.
Notice that the hysteresis sense is inverted between the interval $-90^\circ < \theta < 90^\circ$ to $90^\circ < \theta < 270^\circ$

Some of the magnetization curves also reveal the existence of intermediate smaller steps, the height of which increases at angles close to $\theta =\pm 90^\circ$.
These steps appear typically when magnetization reversal is triggered by a nucleation process as suggested in the previous section.
They arise due to the formation and annhilation of metastable multi-domain magnetic states or due to defects present in the MNP behaving as pinning sites where domain walls remain immobilized up to larger applied magnetic fields.\cite{Wernsdorfer95,Wernsdorfer95b}
The height of each step is related to the total volume of the reversed domain.

The angular dependence of the switching fields is summarized in Fig.~\ref{Fig8}(b) where we plot values of $H_{\rm sw}(\theta)$ at which the first (dots) and last (stars) step is observed (c.f.~vertical dashed lines in Fig.~\ref{Fig8}(a)). 
Experimental data exhibit a clear twofold symmetry along $\theta \sim  8^\circ$ having a small fourfold symmetric contribution at $\theta \sim  98^\circ$. 
This symmetry is highlighted by the black and blue solid lines serving as a guide to the eye.
Such a behavior could reflect the angular dependence of the shape anisotropy (second order) of the particle. 
In addition, magnetization non-uniformities might arise especially at the edges of the particle as a consequence of its shape and edge roughness.
Such non-uniformities behave as nucleation sites for magnetization reversal and might lead to an effective anisotropy of higher degree.\cite{Cowburn00,Han07}
A more complete description of the three-dimensional properties of particle $\#$2 would be possible only by performing magnetization measurements covering any direction in space and is far from the scope of this work.

\section{Conclusions}
\label{sec:conclusions}

A comprehensive characterization of a number of individual cobalt MNPs has been presented.
For this purpose, five different particles having different sizes and aspect ratios have been grown directly on the surface of five ultra-sensitive YBCO-nanoSQUID sensors.
The sensors are based on the use of grain boundary Josephson junctions and have been patterned by FIB milling.
MNPs have been grown by means of FEBID achieving nanometric resolution and, therefore, remarkably large magnetic couplings to the nanoSQUID.

The magnetic volume of each MNP has been estimated from the total magnetic signal sensed by the nanoSQUID and the calculated position-dependent magnetic coupling.
A sizable reduction (by a factor $\sim 3$) of the effective magnetic volume as compared to the geometric one is observed and ascribed to surface oxidation and non-uniform Co concentration in the particle.
The resulting estimated magnetic moments lie within $(1  - 30) \times 10^6\,\mu_{\rm B}$.

Moreover, we have demonstrated that magnetization measurements at magnetic fields $\mu_0 |H| \le 0.15\,$T applied at any direction in the plane of the nanoloop and temperatures $0.3\,{\rm K} < T< 80\,$K are feasible.
Based on these studies, we have distinguished between (quasi) single-domain particles, in which magnetization reversal takes place non-coherently, possibly triggered by a nucleation process, and more complicated topological magnetic states that will be analyzed elsewhere.
Additionally, the energy barriers involved in the reversal process of particle $\#1$ and $\#2$ have been quantified.
Our results demonstrate that YBCO nanoSQUID sensors are outstanding magnetometers well-suited to perform magnetization studies on individual nanomagnets.

\acknowledgments 
We are grateful to J. M. de Teresa for fruitful discussions. M.~J.~M.-P.~acknowledges support by the Alexander von Humboldt Foundation.
This work is supported by the Nachwuchswissenschaftlerprogramm of the Universit\"at T\"ubingen, by the Deutsche Forschungsgemeinschaft (DFG) via Project SFB/TRR 21 C2 and by the EU-FP6-COST Action MP1201.

\bibliography{CoNanoparticles_v15}
\end{document}